\documentclass{article}
\usepackage{spconf,amsmath,graphicx,hyperref, booktabs, subfigure}

\title{Speaker head orientation estimation with a single microphone array using phase spectrogram features  }
\name{Bálint Turi, Archontis Politis, Parthasaarathy Sudarsanam, Tuomas Virtanen}
\address{Audio Research Group, Tampere University, Finland}

\begin{document}
%
\maketitle
\begin{abstract}
Estimating a speaker’s head orientation from audio can provide valuable information in smart environments, meetings, and driver monitoring. We propose a novel approach that leverages the phase component of the short-time Fourier transform from a single microphone array as input to a deep neural network combining convolutional, recurrent, and self-attention layers. Unlike prior methods that use physics-informed handcrafted features or raw waveform inputs, our approach enables robust learning from simulated and real data. Trained on a large-scale dataset generated with voice directivity patterns and fine-tuned on real recordings, our model achieves state-of-the-art accuracy, outperforming baselines under both clean and noisy conditions. Personalization experiments further demonstrate significant gains, reaching a mean angular error of 11.3° when adapting to individual users and environments.

\end{abstract}


\vspace{-3mm}

\section{Introduction}
\label{sec:intro}

\vspace{-1mm}

Estimating a speaker's head orientation has become increasingly important in modern human–machine interaction, as it provides cues about attention, intent, and conversational context. For example, in smart home environments, orientation information can disambiguate which device a user’s command is directed to \cite{Soundr, HOE, DoV}. In meeting rooms, it can help determine who a speaker is talking to, enabling addressee recognition and improved transcription \cite{Ba-meetings, Stiefelhagen-meetings}. In vehicles, it can support driver monitoring systems by tracking attention, contributing to road safety \cite{zhao-car}.

While vision-based estimation of head orientation has been extensively studied \cite{vis-survey}, they require calibrated cameras, are sensitive to occlusions and lighting, and raise privacy concerns. These challenges motivate audio-based solutions that are accurate and more privacy-preserving, leveraging the microphones already embedded in smart speakers, conferencing platforms, and vehicles. Human speech production is inherently directional, as the vocal tract radiates sound unevenly, producing systematic variations in spectral coloration, inter-channel phase differences, and reflection patterns. By exploiting this directional property of speech, audio-only systems can provide a practical and effective means for head orientation estimation.

Early research on acoustic head orientation estimation relied on model-based approaches and large microphone arrays. For example, \cite{sachar2004baseline} employed a Huge Microphone Array (HMA) with 448 microphones, while similar HMA-based systems \cite{Levi-2010, Nakajima-hirofumi-2009} estimated orientation from low-pass filtered signal energy differences. Other studies \cite{alberto-carlos-2006, segura2012gcc} used GCC-PHAT \cite{knapp1976generalized} to extract cross-correlation peaks between microphone pairs, and additional work \cite{segura2008speaker, segura2007multimodal} investigated high-to-low frequency band energy ratios \cite{segura2008speaker}. Although these approaches required minimal training, they were impractical for real-world deployment due to large hardware requirements, computational cost, and sensitivity to noise. Later efforts reduced hardware complexity, with \cite{9287555} using six arrays and \cite{HOE} using two four-channel arrays; however, they still assumed known array geometry and user position. In contrast, single compact devices with small embedded arrays are more aligned with practical applications.

More recently, machine learning approaches have been explored. \cite{Takashima-2011} proposed a hidden Markov model using single-channel audio, though with limited performance compared to multichannel methods. Extensions based on phase coefficients of the cross-power spectrum \cite{Takashima-2012} were adapted from loudspeaker localization and remained noise sensitive. The work in \cite{DoV} predicted eight discrete orientations using a decision tree trained on spectral balance, reverberation, autocorrelation, and GCC-PHAT-based features. Soundr \cite{Soundr} collected 700 minutes of data and trained a CNN+LSTM on raw 16-channel audio. Despite these advances, Soundr reported a mean orientation error of 57° on unseen speakers and rooms, highlighting the ongoing challenge of achieving robust generalization.

Previous approaches have either relied on hand-crafted features based on sound source-to-receiver propagation models~\cite{DoV, speaker-listener, Takashima-2011} or on raw multichannel audio inputs~\cite{Soundr}. Handcrafted features may discard useful information, while raw waveforms require learning representations from scratch, increasing data demands, and overfitting risk.

In this work, we propose a novel approach that uses the \textbf{phase component of the short-time Fourier transform (STFT)} as the primary input feature for a deep neural network that combines convolutional, recurrent, and self-attention layers. Unlike lower-dimensional features or raw audio, our method provides richer directional information without requiring the model to learn feature extraction entirely from limited data. Although STFT is widely used in audio processing, its application to orientation estimation has not been explored, likely due to its high dimensionality and corresponding data requirements. To address this, we pre-train on a large simulated dataset generated by leveraging voice directivity patterns (VDP) before fine-tuning on real recordings, achieving state-of-the-art accuracy and strong generalization across speakers and environments.

\vspace{-2mm}
\section{Problem definition and methodology}
\label{sec:methodology}
\vspace{-1mm}

The objective of this work is to estimate the head orientation of a single speaker in the azimuthal plane within a reverberant room environment, using only a single microphone array with specifications similar to those found in smart home audio devices. In this setting, both the speaker and the array may occupy any position within the room at a fixed height, making the task more general and practical compared to methods that assume fixed or calibrated setups. The angle of orientation $\theta_{ori}$ is defined as the angle between the speaker’s facing direction and the line that connects the speaker's mouth and the center of the microphone array, measured over the full $0^\circ$–$360^\circ$ range. The system takes a single speech utterance as input and outputs one orientation estimate, assuming static head orientation.

\begin{figure}[htb]
    \centering
    \includegraphics[width=0.26\textwidth]{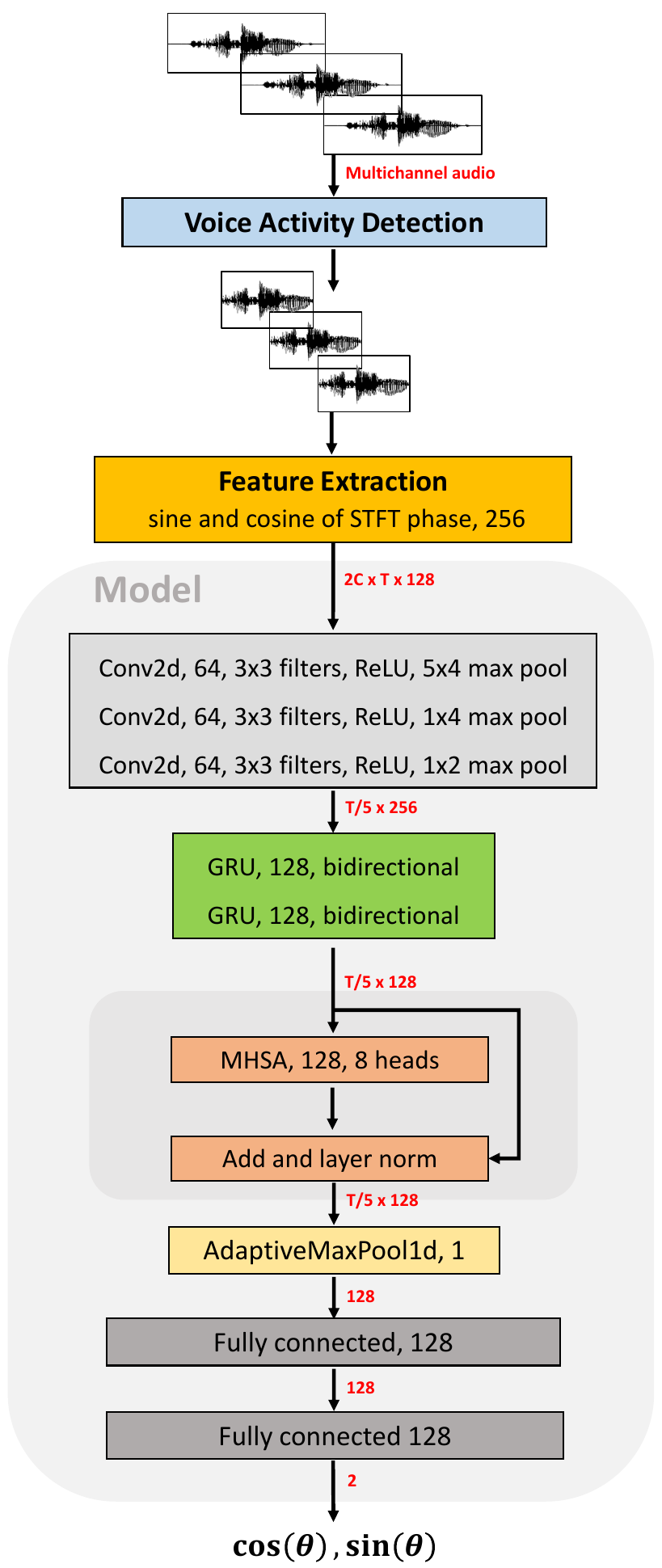}
    \caption{Head orientation estimation pipeline and model architecture overview.}
    \label{fig:pipeline}
\end{figure}

Our proposed framework consists of three main components: speech activity detection, feature extraction, and orientation estimation using a deep neural network. Figure~\ref{fig:pipeline} illustrates the overall pipeline and the model architecture.

\textbf{Voice Activity Detection}:
 Non-speech frames may introduce artifacts that confuse the model. To address this, we apply a state-of-the-art voice activity detection module~\cite{bredin2019pyannoteaudioneuralbuildingblocks} to filter out leading and trailing non-speech segments.

\textbf{Feature Extraction}: For each microphone channel, we compute the STFT using a Hann window with a window size of 4\,ms and a stride of 2\,ms, and then extract the phase. The short window length follows \cite{phase-Peer_2022}, which shows that phase-aware models achieve their best performance in the 2--4\,ms regime. To avoid discontinuities at $\pm \pi$, phase is represented by its sine and cosine values \cite{phase-Peer_2022}. These features are stacked from all channels along the channel dimension, resulting in an input of size $2C \times T \times F$, where $C$ is the number of microphones, $T$ is the number of time frames, and $F = 128$ in our study. Magnitude features, along with phase, were also explored, but we did not find any benefit in using them.

\textbf{Model}:
The model architecture is based on \cite{Adavanne_2019} and \cite{sudarsanam2021assessmentselfattentionlearnedfeatures}. It accepts multichannel phase features of size $2C \times T \times 128$ as input and processes them through three 2D convolutional layers with $3 \times 3$ kernels, 64 filters, and ReLU activations, each followed by max pooling with kernel sizes $5 \times 4$, $1 \times 4$, and $1 \times 2$, reducing the temporal and frequency resolutions while preserving spatial information. Reshaping produces feature maps of size $T/5 \times 256$, which are passed through two bidirectional GRU layers with 128 hidden units to capture temporal dependencies, producing embeddings of size $T/5 \times 128$. These embeddings are further processed by two blocks of multi-head self-attention with 8 heads and an attention size of 128, with residual connections and layer normalization. To handle variable-length signals, the resulting features are aggregated by an adaptive max pooling layer to size 128, which is passed through a fully connected layer followed by a linear projection producing the orientation representation $\bigl[\cos(\hat{\theta}), \sin(\hat{\theta})\bigr]$, where $\hat{\theta}$ denotes the predicted orientation angle.
Predicting sine and cosine avoids $0^\circ/360^\circ$ discontinuities. The final orientation is recovered as
\vspace{-1mm}
\[
\hat{\theta} = \big(\operatorname{atan2}(\sin(\hat{\theta}), \cos(\hat{\theta})) \times \tfrac{180}{\pi}\big) \bmod 360 .
\]

\vspace{-5mm}
\section{Evaluation}
\label{sec:evaluation}
\vspace{-2mm}

We evaluate our proposed method on both simulated and real datasets. The simulated setup allows controlled testing across acoustic conditions and baselines. We then examine the effect of personalization through user- and room-specific fine-tuning, and finally validate generalization on real recordings. Performance is measured using classification accuracy for discrete angle tasks and mean angular error (MAE), defined as
\vspace{-2mm}
\[
\text{MAE} = \tfrac{1}{N} \sum_{i=1}^{N} 
\min\!\big( \left| \theta_{i} - \hat{\theta_{i}} \right|, \; 360^{\circ} - \left| \theta_{i} - \hat{\theta_{i}} \right| \big),
\] for continuous predictions. Here, $\theta_{i}$ and $\hat{\theta_{i}}$ denote the true and predicted angles for sample $i$, and $N$ is the total number of samples.

\vspace{-2mm}

\subsection{Simulated Dataset Generation}

We generate a simulated dataset using 22 voice directivity patterns from \cite{dirpat-Brandner2018, sopranom-Shabtai2017, usyd-Cabrera2011} to simulate a human speaker as a sound source, in combination with speech samples from the VCTK corpus \cite{vctk-Yamagishi2019}. For each utterance, we randomly select (i) a VDP, (ii) an orientation angle uniformly sampled from $0^{\circ}$–$360^{\circ}$, and (iii) a room configuration, with room dimensions drawn uniformly from $[3, 12] \times [3, 12] \times [2, 6]$ meters and wall absorption and scattering coefficients sampled from $(0.2, 0.6)$ and $(0.1, 0.5)$, respectively. The speaker and microphone array positions are sampled uniformly within the room and placed at a fixed height of 1.5\,m in a full 3D simulation. The array is modeled as a compact circular array with a radius of 4.5\,cm and six microphones evenly spaced at $60^{\circ}$ intervals, reflecting the type of hardware commonly used in smart audio devices. Acoustic propagation is simulated using the image source method in Pyroomacoustics \cite{pyroomacoustics-Scheibler_2018} with the reflection order limited to 20, resulting in a mean RT$_{60}$ of approximately 0.85\,s with variance 0.07\,s$^2$, and each utterance is truncated to a maximum duration of 3\,s. VCTK provides 44k recordings from 110 speakers. To test true generalization, eleven speakers and six VDPs are held out for testing, yielding 40,295 training and 3,947 test utterances drawn from the same distribution.

\subsection{Baselines}

We compare our method against three state-of-the-art approaches. The first is Soundr \cite{Soundr}, a CNN+LSTM model that takes raw 16-channel audio as input. For consistency with our experimental setup, we re-implement their CNN encoder, replace our CNN layers with it, and evaluate it under our simulation framework. The second is \cite{speaker-listener}, which is based on interaural level differences (ILD), interaural time differences (ITD), high-low-frequency band separation, and inverse convolution features. We replicate their feature extraction pipeline and provide these features as input to our network. The third is \cite{DoV}, which uses a set of features informed by the physics of speech propagation. Specifically, they compute spectral balance features (low- and high-frequency power, their ratio, polynomial and linear regression coefficients over the FFT), wavefront crispness measures (autocorrelation ratios, statistical descriptors, and speech-to-reverberation modulation energy ratio), and cross-microphone correlation features using GCC-PHAT (raw correlations, peak values, interchannel delays, area under the curve, and time-difference-of-arrival statistics). Since their dataset is public, instead of re-implementing their method in our environment, we directly evaluate our approach on their real dataset.

All models are optimized using mean squared error (MSE) loss and trained with the Adam optimizer~\cite{kingma2017adammethodstochasticoptimization}, starting from an initial learning rate of $4 \times 10^{-4}$ that decays linearly. On the simulated dataset, each configuration is trained for 200,000 iterations with a batch size of 16.

\vspace{-2mm}
\subsection{Results on Simulated Data}

To examine robustness in realistic acoustic environments, we augment both the training and testing subsets of the simulated dataset with ambient noise samples from the WHAM dataset \cite{Wichern2019WHAM}. Since the noise dataset is monophonic, we generate a realistic multichannel version by phase-randomizing uncorrelated copies and mixing them to match the target inter-channel coherences under isotropic diffuse field assumptions, following \cite{mccormack2021rendering, mirabilii2021generating}. In addition to three noise conditions (clean, moderate noise at 10–20 dB SNR, and high noise at 0–10 dB SNR), we also evaluate an anechoic clean setup to examine the role of room reverberation. 

Table~\ref{tab:sim-results} summarizes the results for all methods and noise levels. STFT phase significantly outperforms the baselines across all noisy and reverberant conditions. Moreover, both raw audio and STFT phase perform worse without reverberation, whereas ITD \& ILD benefit from the anechoic condition. Overall, moderate reverberation appears beneficial for phase- and raw waveform-based features but harmful for ITD \& ILD, indicating that using phase features of the STFT provides higher robustness to environmental interference compared to raw audio or handcrafted cues.

\begin{table}[h!]
\centering
\caption{Comparison of orientation estimation methods in different acoustic conditions, measured in MAE . Noise levels indicate SNR: 10–20 dB (moderate) and 0–10 dB (high).}
\label{tab:sim-results}
\resizebox{\columnwidth}{!}{%
\begin{tabular}{lcccc}
\toprule
\textbf{Method} & \textbf{Anechoic} & \textbf{Clean} & \textbf{10–20 dB} & \textbf{0–10 dB} \\
\midrule
Raw audio + CNN \cite{Soundr} & 56.9° & 44.8° & 63.7° & 75.1° \\
ITD \& ILD \cite{speaker-listener} & 47.7° & 52.7° & 74.4° & 82.8° \\
\textbf{Phase of STFT} & \textbf{28.4°} & \textbf{19.9°} & \textbf{25.6°} & \textbf{29.5°} \\
\bottomrule
\end{tabular}}
\end{table}

\begin{figure}[t]
    \centering
    \subfigure[STFT Phase Anechoic]{
        \includegraphics[width=0.47\linewidth]{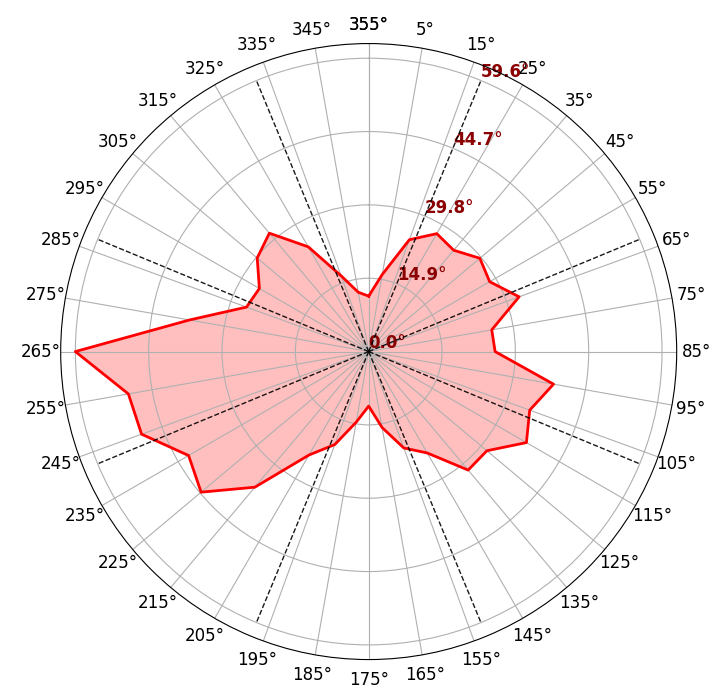}
    }
    \hfill
    \subfigure[STFT Phase Reverberant]{
        \includegraphics[width=0.47\linewidth]{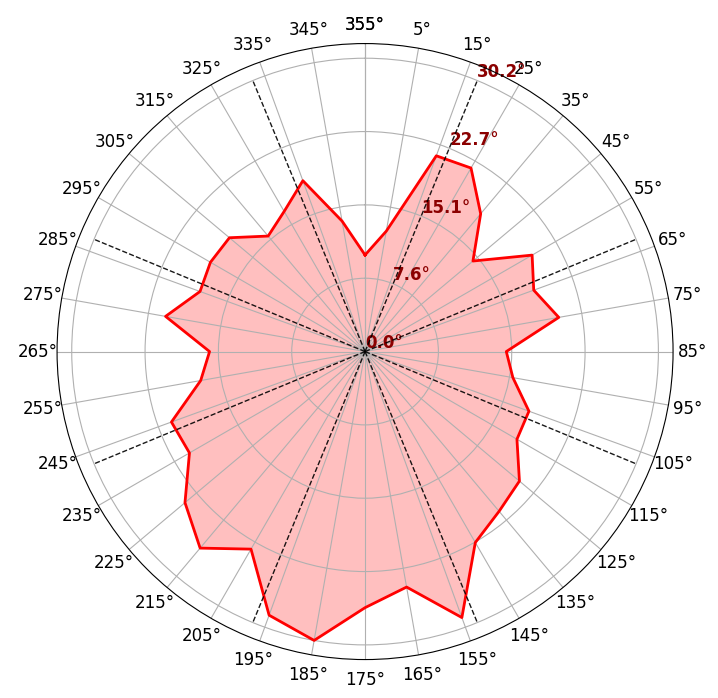}
    }
    \caption{Mean angular error in 10$^\circ$ bins for the STFT-phase model under anechoic and reverberant conditions.}
    \label{fig:spider}
\end{figure}

Figure~\ref{fig:spider} shows the mean angular error aggregated in 10$^\circ$ bins. 
In anechoic conditions, errors are lowest at 0$^\circ$ and 180$^\circ$ but increase 
substantially for lateral orientations ($\pm90^\circ$), indicating ambiguity in 
phase cues without reflections. In contrast, reverberation produces a more uniform error distribution across azimuths, suggesting that early reflections 
introduce additional orientation-dependent phase structure that the model exploits.

In many real-world applications, users may be asked to provide a small number of samples to calibrate the model, since machine learning models are expected to work better on matched data from the speakers and environments they were trained on. To investigate this effect, we evaluate our method under three personalization scenarios.

\emph{Room Only} case: The model is fine-tuned using samples from the target environment but tested on unseen speakers. For this setup, we created ten fixed rooms and used 500 samples per room for fine-tuning, then evaluated on 400 samples per speaker from ten unseen speakers in these rooms.

\emph{Speaker Only} case: The model is adapted using samples from the target speaker but tested in unseen rooms. To simulate this, we selected ten unseen speakers and used 150 samples per speaker for fine-tuning and 250 for testing.

\emph{Speaker + Room} case: The model is fine-tuned using samples from both the target speaker and the target room, neither of which the model has encountered during pre-training. For this, we created a fixed room for each speaker and used 150 samples per speaker for fine-tuning and 250 for testing.

This allows us to measure how much performance can be gained from personalization to the user and/or adaptation to the environment. As shown in Table~\ref{tab:room-user-results}, personalization consistently improves accuracy, with the largest gains observed when both user and room information are available. Notably, adapting to users provides a larger benefit than adapting to rooms, highlighting the strong variability in speech directivity between speakers.

\begin{table}[h!]
\centering
\caption{MAE across different personalization scenarios.}
\label{tab:room-user-results}
\resizebox{0.45\columnwidth}{!}{%
\begin{tabular}{l c}
\toprule
\textbf{Prior knowledge} & \textbf{MAE} \\
\midrule
None (Baseline) & 19.9° \\
Room Only & 17.6° \\
Speaker Only & 14.2° \\
Speaker + Room & \textbf{11.3°} \\
\bottomrule
\end{tabular}}
\end{table}

\vspace{-3mm}

\subsection{Results on Real Data}

To further validate our approach, we evaluate on the real dataset from \cite{DoV}, which contains samples in eight discrete azimuthal orientations. Following their experimental protocol, we replicate the ``Per Angle Classifier'' setup by replacing the final regression layer with an 8-class linear classifier layer. We use the provided split to fine-tune the classifier on all session one samples, and to test it on session two samples, and vice versa. As shown in Table \ref{tab:real-results}, our method outperforms the \cite{DoV} baseline, while training only on real data performs worse, highlighting the benefit of pre-training on large-scale simulated data.

\begin{table}[h!]
\centering
\caption{Accuracy on the real dataset \cite{DoV} under different training strategies.}
\label{tab:real-results}
\resizebox{0.95\columnwidth}{!}{%
\begin{tabular}{l c}
\toprule
\textbf{Method} & \textbf{Accuracy} \\
\midrule
DoV baseline \cite{DoV} & 65.4\% \\
Ours (trained only on real data) & 62.6\% \\
Ours (pre-trained on simulated + fine-tuned on real) & \textbf{73.2\%} \\
\bottomrule
\end{tabular}}
\end{table}

\vspace{-4mm}

\subsection{Discussion}

The experimental results highlight several key insights. First, the strong performance on both simulated and real datasets demonstrates that STFT phase features are superior to raw audio and hand-crafted inter-channel features, offering greater robustness to noise where other baseline methods degrade substantially.  We hypothesize that the phase provides reliable cues for the relative rotation between the source and microphone array, including echoes from image sources, which the model can exploit. Similar importance of phase features was observed in DNN-based speaker distance estimation with a single microphone \cite{neri2024speaker}, though further experiments are needed to validate these hypotheses.

Second, the personalization experiments confirm that user-specific information contributes more to accuracy improvements than environment-specific information, reflecting the variability of speech radiation patterns across individuals. Fine-tuning on user data can yield significant gains.

Third, the evaluation on the real dataset from \cite{DoV} further validates the generalization capability of our method and the effectiveness of pre-training on large-scale simulated data, providing a practical way to mitigate limited annotated real-world data.

Finally, while our method achieves state-of-the-art results, several limitations remain. The personalization results reveal that the model still lacks strong generalization, as performance improves significantly when adapted to new users or environments. This suggests that more diverse and larger datasets are required to achieve robust generalization and higher accuracy. 

Overall, these findings demonstrate that leveraging STFT phase-based features within a deep neural architecture offers a promising and scalable path toward accurate and noise-robust speaker head orientation estimation. 

\vspace{-4mm}

\section{Conclusion}
\label{sec:conclusion}

\vspace{-2mm}

We presented a novel approach to speaker head orientation estimation that leverages phase features from the short-time Fourier transform in combination with a deep neural network. Through experiments on both simulated and real datasets, we demonstrated that this representation provides superior robustness to noise and generalization across speakers and environments compared to existing baselines. Furthermore, our results show that fine-tuning the model on even a small amount of user- or environment-specific data can significantly improve accuracy, highlighting the value of personalization in practical deployment scenarios.


\vfill\pagebreak

\small
\bibliographystyle{IEEEbib}
\bibliography{strings,refs}

\end{document}